\documentstyle[aps,prl,epsfig]{revtex}

\begin{document}
\twocolumn[\hsize\textwidth\columnwidth\hsize\csname@twocolumnfalse\endcsname
\draft

\title{Density of states modulation in the pseudogap state of high-$T_c$ superconductors}

\author{Xin-Zhong Yan}

\address{Institute of Physics, Chinese Academy of Sciences, P.O. Box603, Beijing 100080, China\\}

\date{\today}

\maketitle

\widetext
\begin{abstract}
The density of states modulation recently observed by the scanning tunneling microscopic experiment in the pseudogap state of high-$T_c$ superconductors is explained by the pairing assisted particle transitions under perturbation of a periodic pairing interaction. In such a transition process, a particle with momentum ${\bf k}$ firstly picks up another particle of inverse spin to produce a pair leaving a hole. Under the perturbation of periodic pairing interaction of modulation wave vector ${\bf Q}$, the pair absorbs a momentum ${\bf Q}$ and then breaks into two single-particles: one propagates with momentum ${\bf k+Q}$, and another one fills in the hole. The transition is significant at low energies within the pseudogap since where the pairing excitations most favorably survive. We calculate the Fourier component of the modulated density of states using two different models. Both of the theoretical results are consistent with the experiment.

\end{abstract}


\vfill
\narrowtext

\vskip2pc] 

The recent scanning tunneling microscopic (STM) experiments on underdoped Bi$_2$Sr$_2$CaCu$_2$O$_{8+\delta}$ samples have detected a dispersionless spatial modulation close to the periodicity of $4a\times4a$ in the tunneling conductance in the pseudogap states\cite{Vershinin}. This modulation seems to be different from that in the superconducting state, which exhibits energy dispersion\cite{Howald,Hoffman1,McElroy}. The STM observations show a distinct enhancement of intensity in the modulated patterns within the pseudogap energy scales. A similar modulation has been observed in the vortex cores\cite{Hoffman2}. This phenomenon provides a new clue for understanding the physics of the pseudogap states.  

The mechanism of modulated electronic structures has been studied by various models of electron ordering\cite{Polkovnikov,Chen1,Franz,Zhu,Podolsky,Wang,Sachdev,Chen2}. The possible origin of the modulation in the superconducting state is attributed to the quantum interferences of quasiparticle scattering by some kind of perturbation\cite{Wang}. For explaining the dispersionless modulation in the pseudogap states, Chen {\it et al.} have proposed a model of density wave of $d$-wave Cooper pairs without global phase coherence\cite{Chen2}. The Fourier component of the tunneling density of state (DOS) $\rho({\bf Q},E)$ is predicted to be approximately even with respect to the energy $E$. This result captures the main feature of the energy dependence of the experimental observations. They noted that in case of charge density wave (CDW) or potential scattering, $\rho({\bf Q},E)$ is approximately antisymmetric with $E \to -E$. Very recently, Anderson has suggested a model of Wigner crystal of $d$-wave hole pairs embedded in a background of $d$-wave resonance valence bond of singlet electron pairs and argued this structure of holes might be also relevant with the anomalous ${1\over 8}$-doping phenomenon\cite{Anderson}.

In this paper, we will propose a description that the modulation phenomenon in the pseudogap state is a consequence of pairing assisted single-particle transitions under the perturbation of the modulated pairing interaction. Instead of supposing a crystal of pairs, we consider the pseudogap state as a system of single-electrons coexisting with the uncondensed pairs. Such a description for the pseudogap state is consistent with the existed models for studying the superconducting phase transition and the transport properties\cite{Dahm,Levin,Daggoto,Yanase,Yan1,Yan2}. By model calculation of the Fourier component of the modulated DOS that essentially equivalent to the modulated tunneling conductance at low temperatures, we will show the present model can reasonably explain the STM experimental result. 

\vskip -4mm
\begin{figure}
\centerline{\epsfig{file=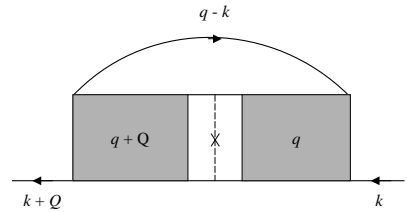,width=6 cm}}
\vskip -20mm
\caption{First order self-energy of the Green's function. The dashed line with a cross denotes the pairing interaction with modulation wave number ${\bf Q}$. The shaded parts represent the ladder diagrams.}
\end{figure}

The process of pairing assisted single-particle transition is illustrated by the self-energy of the first order perturbation as in Fig. 1, where the dashed line with a cross denotes the modulated pairing interaction of modulation wave number ${\bf Q}$, and shaded rectangles are the ladder diagrams representing in essence the pair propagation. The quantities $k = ({\bf k}, z_n)$ and $q = ({\bf q}, Z_m)$ are the generalized momenta with $z_n$ and $Z_m$ the imaginary Matsubara frequencies respectively of fermions and bosons. $q+Q$ is understood as $({\bf q+Q}, Z_m)$. In the process, a particle of momentum $k$ firstly picks up another particle of inverse spin with momentum $q-k$ to make up a singlet pair of total momentum $q$. Under the perturbation of the periodic pairing interaction, the pair transits to a state of momentum $q+Q$. When it breaks, it emits two single-particles: one fills in the hole, and another one propagates with momentum $k+Q$. As a result, the single-particle jumps from $k$ to $k+Q$ through the assistance of the pair transition. Since the pair excitations are pronouncedly only within the pseudogap, such a single-particle transition and thereby the DOS modulation are significant within the same energy scales. In the following, we firstly illustrate that the DOS modulation is approximately even with $E \to -E$, and then in the later part of this paper, we present our model calculations.

At temperature $T$ close to the transition point $T_c$, the pair propagator has strong peak at $q = 0$, which implies the pair excitations happen mostly at low energies\cite{Levin,Yan1,Yan2}. Especially this peak diverges at $T = T_c$. To qualitatively analyze the DOS modulation, we here firstly consider the extreme case that the contribution to the first order self-energy $\Sigma^{(1)}(k,Q)$ predominantly comes from small regions around $q = 0$ and $q+Q = 0$ (at which the right and left pair propagators respectively in Fig. 1 have strong peaks). We then have approximately $\Sigma^{(1)}(k,Q) \propto [G(-k)+G(-k-Q)]f({\bf k,Q})$ with $f({\bf k,Q})$ an interference factor [see Eq. (\ref{PSE})]. The first order Green's function behaves like $G^{(1)}(k,Q)\propto G(k+Q)[G(-k)+G(-k-Q)]G(k)f({\bf k,Q})$. The Fourier component of the modulated DOS is defined by
\begin{equation}
\rho ({\bf Q},E)= -{1\over \pi N}\sum_{\bf k} {\rm Im}G^{(1)}({\bf k,Q}, E+i0^+), \label{MDOS}
\end{equation}
where $N$ is the total number of sites in a lattice on which the electron system is modeled. By the approximation above, we have
\begin{equation}
\begin{array}{rl}
\rho ({\bf Q},E) \propto &\sum\limits_{\bf k}{\rm Im}[G^R({\bf k+Q},E)G^A({\bf -k},-E)G^R({\bf k},E)\cr\cr
&~~~~~~+({\bf k}\leftrightarrow{\bf k+Q})]f({\bf k,Q}).
\end{array} \label{AMDOS}
\end{equation}
The ${\bf k}$-summation comes predominately from the regions close to the quasiparticle poles where the Green's function behaves like
\begin{equation}
G^R({\bf k},E)= [G^A({\bf k},E)]^\ast \approx {a_{\bf k}\over E-E_{\bf k}+i\gamma_{\bf k}}, \label{Gf}
\end{equation}
where $E_{\bf k}$ and $\gamma_{\bf k}$ are respectively the energy and width of a quasiparticle, and $a_{\bf k}$ is the residue at the pole. At low energy $E \sim 0$, only those states of $E_{\bf k} \approx E_{\bf k+Q} \sim E$ in some regions close to the Fermi surface are important to the ${\bf k}$-summation. For example, ${\bf Q} = (\pi/2, 0)$ (in unit of lattice constant $a = 1$), these regions are near to $(0, \pm\pi)$. When $E$ is changed to $-E$, the corresponding states are changed to another side of the Fermi surface. Under the change $G^R({\bf k},E)\approx -G^A({\bf k'},-E)$ with $E_{\bf k'}=-E_{\bf k}$, a minus sign comes from the three Green's functions in the square bracket in Eq. (\ref{AMDOS}). This minus sign is just canceled by the imaginary part of the complex conjugate. On the other hand, since $f({\bf k',Q})$ should not be different too much from $f({\bf k,Q})$ and the unperturbed densities of those important states of respectively $k$ and $k'$ are approximately the same, we thus have $\rho ({\bf Q},E) \approx \rho ({\bf Q},-E)$. From the discussion, we can draw an outline for $\rho ({\bf Q},E)$ at small $E$: there is an approximately symmetric peak at $E = 0$ with width of the pseudogap energy scale.

To present our quantitative calculation, we start with the perturbation of the modulated pairing interaction, 
\begin{equation}
H' = \sum_{ij} V_{ij}n_{i\uparrow}n_{j\downarrow}, \label{pert}
\end{equation}
with $V_{ij} = V_0(|{\bf r_i-r_j}|)\exp[{\bf Q}\cdot({\bf r_i+r_j})/2]$ and $n_{i\alpha}$ the density operator of electron with spin-$\alpha$ at site $i$. By first order perturbation, since each Fourier component of the DOS modulation can be analyzed separately, we consider a single modulation wave vector and set ${\bf Q} = (\pi/2,0)$ in the present calculation. We will consider the singlet $d$-wave pairing of electrons at the nearest-neighbor sites. It is then convenient to write the perturbation Hamiltonian in ${\bf k}$-space,
\begin{equation}
H' = {v_0\over 2N}\sum_{\bf q} (p^{\dagger}_{\bf q+Q}p_{\bf q}+p^{\dagger}_{\bf q}p_{\bf q+Q}), \label{ksp}
\end{equation}
where $v_0$ is the interaction strength of $d$-wave channel, and $p_{\bf q} = \sum\limits_{\bf k}\eta_{\bf k}c_{{\bf k+q/2}\uparrow}c_{{\bf -k+q/2}\downarrow}$ is the $d$-wave pair operator, with $c_{{\bf k}\alpha}^{\dagger }$ ($c_{{\bf k}\alpha}$) the creation (annihilation) operator for electrons with momentum-${\bf k}$ and spin-$\alpha $ and $\eta_{\bf k} = \cos k_x -\cos k_y$.

For the unperturbed electron system, we firstly consider a tight-binding model with $d$-wave channel interaction\cite{Levin,Daggoto,Yan1}. The Hamiltonian is given by
\begin{equation}
H = \sum_{{\bf k}\alpha}\xi_{\bf k} c_{{\bf k}\alpha }^{\dagger }c_{{\bf k}\alpha }-
{\frac{v}{N}}\sum_{\bf q}p^{\dagger }_{\bf q}p_{\bf q} \label{H1}
\end{equation}
where $\xi_{\bf k} = -2t(\cos k_x + \cos k_y) - 2t_z\cos k_z -\mu$ with $\mu$ the chemical potential. For taking into account of the effect of short-range strong Coulomb repulsion, the hopping integrals $t$ and $t_z$ are considered as to be proportional to the doping concentration $\delta$, e.g., $t = t_0\delta$ with $t_0$ a constant. For the quasi-two-dimensional system, $t_z/t = 0.01$ is supposed. The weak interlayer coupling ensures a finite transition temperature\cite{Levin,Yan1}. In the present calculation, we take $v/t_0 = 0.2$ and $\delta = 0.125$. Throughout this paper, we use the units in which $\hbar = k_B = 1$.
 
The single-particle Green's function for the unperturbed pseudogap state reads
\begin{equation}
G({\bf k},z_n) = [z_n - \xi_{\bf k} - \Sigma({\bf k},z_n)]^{-1} \label{GRF}
\end{equation}
with $\Sigma({\bf k},z_n)$ the self-energy. For brevity, we will occasionally use the generalized momentum defined earlier, e.g., $G({\bf k},z_n)$ is simply written as $G(k)$. Within the model given by Eq. (\ref{H1}), the most important effect on the single-particle comes from the pair excitations. In the self-energy, we take into account this pairing effect through the ladder-diagram that is a conserving approximation. The self-energy is approximated as\cite{Yan1}
\begin{equation}
\Sigma(k) = {\frac{T}{N}}\sum_{q}\eta^2_{\bf k-q/2}G(q-k)P(q), \label{SE}
\end{equation}
where $P(q)$ is the summation result of the ladder-diagram, and is given by
\begin{eqnarray}
P(q) &=& \frac{v^2\Pi(q)}{1 + v\Pi(q} \label{PF} \\
\Pi(q) &=& -{\frac{T}{N}}\sum_{{\bf k}n}\eta^2_{\bf k}G({\bf k}^+,z_n) G({\bf k}^-,Z_m-z_n) \label{PiF}
\end{eqnarray}
with ${\bf k}^{\pm} = {\bf k}\pm {\bf q/2}$. With the computation algorithms recently developed in Ref. \onlinecite{Yan2}, we have obtained a self-consistent solution to the Green's function.

With the result for the unperturbed Green's function, the first order self-energy $\Sigma^{(1)}(k,{\bf Q})$ can be obtained immediately. For the convenience of numerical calculation, we express $\Sigma^{(1)}(k,{\bf Q})$ in the following form by denoting in Fig. 1 the input and output momenta with ${\bf k-Q/2}$ and ${\bf k+Q/2}$, respectively,
\begin{equation}
\Sigma^{(1)}(k,{\bf Q}) = {\frac{v_0}{v^2}}{\frac{T}{N}}\sum_{q}f({\bf k,q,Q})G(q-k)P(q+Q/2)P(q-Q/2) \label{PSE}
\end{equation}
with $f({\bf k,q,Q})=\eta_{\bf k+Q/4-q/2}\eta_{\bf k-Q/4-q/2}$. By noting that $\eta_{\bf k-q/2}$ can be factorized as $\eta_{\bf k-q/2}= \psi^{\dagger}({\bf k})\varphi({\bf q})$ with $\psi^{\dagger}({\bf k})=(\cos k_x,\sin k_x,-\cos k_y,-\sin k_y)$ and $\varphi^{\dagger} ({\bf q})=[\cos (q_x/2),\sin (q_x/2),\cos (q_y/2),\sin (q_y/2)]$, the ${\bf q}$-summation in Eq. (\ref{PSE}) can be manipulated by fast Fourier transform. The expression for the first order Green's function is given by
\begin{equation}
G^{(1)}(k,{\bf Q}) = G(k+Q/2)\Sigma^{(1)}(k,{\bf Q})G(k-Q/2). \label{G1}
\end{equation}
By firstly integrating $G^{(1)}(k,{\bf Q})$ with respect to the momentum ${\bf k}$ and then using the Pad\'e approximation for analytical continuation of the result from the imaginary Matsubara frequency to the real energy\cite{Vidberg}, we then obtain the result for $\rho ({\bf Q},E)$. 

Shown in Fig. 2 are results for $\rho ({\bf Q},E)$ at temperatures $T/t_0 = 0.0175$, 0.025, and 0.0375 all above the transition point. [The values of $-v\Pi(0)$ are 0.925, 0.827, and 0.700, respectively. The transition point corresponds to $-v\Pi(0)=1$.] As expected, there is a central peak in $\rho ({\bf Q},E)$ nearly symmetric about $E = 0$. This is consistent with the experiment\cite{Vershinin}. Out of the pseudogap region, $\rho ({\bf Q},E)$ is negative, which should be consistent with a sum rule\cite{Podolsky}. 

Though the calculated $\rho ({\bf Q},E)$ can reasonably reflect the feature of the experimental observation, the phenomenological model seems to be too simple. The model cannot correctly treat the short-range antiferromagnetic correlation between the electrons. Because of this reason, we have considered a more realistic model, the quasi-two dimensional Hubbard model. However, we will see that the results of the two models have the common feature as shown in Fig. 2; the qualitative behavior of the DOS modulation predicted by the process shown in Fig. 1 insensitively depends on the models.

\vskip -4mm
\begin{figure}
\centerline{\epsfig{file=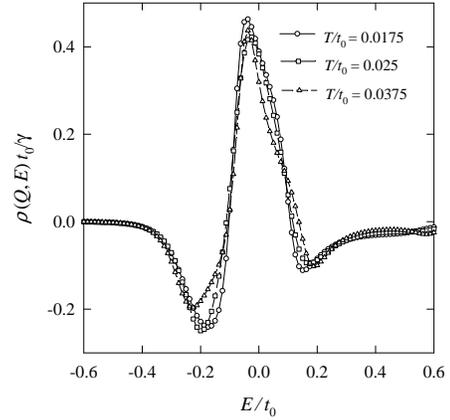,width=7.5 cm}}
\vskip -4mm
\caption{$\rho ({\bf Q},E)$ as function of $E$ at various temperatures. The dependence on the perturbation strength $v_0$ is eliminated by dividing the constant $\gamma=v_0/v$.}
\end{figure}

For the unperturbed state of the Hubbard model, we have recently obtained self-consistent solution using the charge, spin and pairing-fluctuation exchange approximation\cite{Yan2}. By this approximation, besides the pair excitations are taken into account by the ladder-diagram, the effect of short-range Coulomb repulsion is treated by the standard charge and spin-fluctuation exchange. In case of the Hubbard model, the functions $P(q\pm Q/2)$ appeared in Eq. (\ref{PSE}) are modified to be $P(q\pm Q/2)= v^2\Pi_0(q\pm Q/2)/[1 + v\Pi(q\pm Q/2)]$ with
\begin{eqnarray}
\Pi_0(q) &=& -{\frac{T}{N}}\sum_{k}\eta_{\bf k-q/2}\phi(k)G(k) G(q-k) \label{HPiF0}\\
\Pi(q) &=& -{\frac{T}{N}}\sum_{k}\phi^2(k)G(k) G(q-k) \label{HPiF}
\end{eqnarray}
where $\phi(k)$ is the pair wave function of $d$-wave symmetry. Under a normalization condition for $\phi(k)$, the coupling strength $v$ in this case is obtained as $v = \lambda/2$ with $\lambda$ the eigenvalue of the \'Eliashberg equation\cite{Yan2}. Correspondingly, the interference factor $f({\bf k,q,Q})$ in Eq. (\ref{PSE}) should be changed to $f({\bf k,q,Q})=\phi(k+Q/4-q/2)\phi(k-Q/4-q/2)$. However, with such an interference factor, it is nearly impossible to get a numerical result for $\Sigma^{(1)}(k,{\bf Q})$; the amount of the computation is tremendous. We then make an approximation by factorizing $\phi(k\pm Q/4-q/2) \propto \Delta(z_n)\eta_{\bf k\pm Q/4-q/2}$ with $\Delta(z_n)= \phi({\bf k_M},z_n)/2$ and ${\bf k_M}=(\pi,0)$. That means the wave function of a pair has approximately the nearest-neighbor $d$-wave pairing structure and its frequency dependence is separated by the factor $\Delta(z_n)$. The proportional coefficient is determined by the normalization condition
\begin{equation}
{\frac{v_pT}{N}}\sum\limits_{k}\Delta^2(z_n)\eta^2_{\bf k}G(k) G(-k) = \lambda
\end{equation}
by which an effective coupling constant $v_p$ is so defined. 

\vskip -4mm
\begin{figure}
\centerline{\epsfig{file=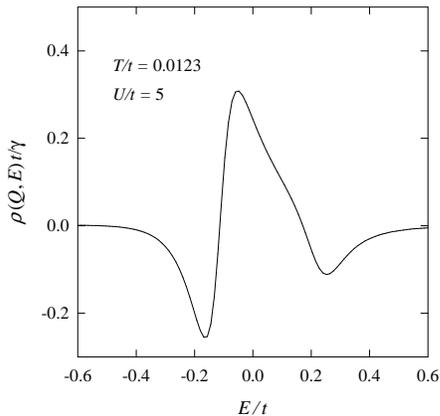,width=7.5 cm}}
\vskip -4mm
\caption{$\rho ({\bf Q},E)$ as function of $E$ for the Hubbard model. The constant $\gamma$ is given by $v_0/v_p$.}
\end{figure}

In Fig. 3, we depict the result for $\rho ({\bf Q},E)$ for the Hubbard model at $U = 5t$, $\delta = 0.125$ and $T = 0.0123t$, where $\lambda$ = 0.982 and $v_p = 15.5$ are obtained. The main feature of this result is approximately the same as in Fig. 2; both of them show a central peak at $E$ = 0. From Fig. 3, the width of the peak is seen about $0.2t$ that is overall the same as the pseudogap width found in the unperturbed DOS\cite{Yan2}. This is again consistent with the experiment. The central peak seems less symmetric than expected. This is because the unperturbed DOS of the Hubbard model by the approximation is not symmetric. Near the Fermi energy, the unperturbed DOS at $E < 0$ is larger than that at $E>0$, which means the summation in Eq. (1) over the states below the Fermi surface is larger and results in a shift of the maximum of $\rho ({\bf Q},E)$ toward to a slightly negative energy.  

In summary, we have shown that the DOS modulation observed in the pseudogap state of high-temperature superconductors can be explained by the pairing assisted particle transitions under the perturbation of modulated pairing interaction. Such a transition process is illustrated in Fig. 1. The calculated Fourier component $\rho ({\bf Q},E)$ as a function of energy has a central peak at $E = 0$ with a width of pseudogap energy scale in consistent with the experiment.   

The author thanks Prof. C. S. Ting for useful discussions on the related problems. This work was supported by Natural Science Foundation of China under grant number 10174092 and by Department of Science and Technology of China under grant number G1999064509.


\end{document}